\documentclass[ aps, showpacs, showkeys, nofootinbib, floatfix, superscriptaddress]{revtex4}

\usepackage{amsfonts}

\usepackage{amssymb}
\usepackage{amsmath}
\usepackage{graphicx}

\begin{document}

\title{Cosmology in symmetric teleparallel gravity and its dynamical system}
 \author{Jianbo Lu}
 \email{lvjianbo819@163.com}
 \affiliation{Department of Physics, Liaoning Normal University, Dalian 116029, P. R. China}
 \author{Xin Zhao}
 \affiliation{Department of Physics, Liaoning Normal University, Dalian 116029, P. R. China}
 \author{Guoying Chee}
 \affiliation{Department of Physics, Liaoning Normal University, Dalian 116029, P. R. China}
 \email{qgy8475@sina.com}

\begin{abstract}
 We explore an extension of the symmetric teleparallel gravity denoted the $f(Q)$ theory, by considering a function of the nonmetricity invariant $Q$ as
the gravitational Lagrangian. Some interesting properties could be found in the $f(Q)$ theory by comparing with the  $f(R)$ and $f(T)$ theories. The field equations are derived in the $f(Q)$ theory. The cosmological application is investigated. In this theory the accelerating expansion is an intrinsic property of the universe geometry without need of either exotic dark energy or extra fields. And  the state equation of the geometrical dark energy can cross over the phantom divide line in the $f(Q)$ theory. In addition, the dynamical system method are investigated. It is shown that there are five critical points in the STG model for taking $f(Q)=Q+\alpha Q^2$. The critical points $P_{4}$ and $P_{5}$ are stable.  $P_{4}$ corresponds to the geometrical dark energy dominated de Sitter universe ($w_{tot}^{eff}$=-1), while $P_{5}$ corresponds to the matter dominated universe ($w_{tot}^{eff}$=0). Given that $P_{4}$ represents an attractor, the  cosmological constant problems, such as the fine tuning problem, could be solved in the STG model.

\end{abstract}



\maketitle

\section{$\text{Introduction}$}

 The discovery of the cosmic accelerated expansion has motivated a vast number of researches on modifications of general relativity (GR) (for recent reviews, see [1-3]) to explain this acceleration. A plethora of theories have been proposed in the literature, essentially based on specific approaches. From the view point of the connection, theories of gravity can be classified into three broad classes. The first class uses the Levi-Civita connection of the metric and its curvature. The second class uses the tetrads of a metric and their curvature free, metric-compatible, Weitzenb$\ddot{o}$ck connection with torsion. And the third class uses a curvature-free and torsion-free symmetric teleparallel connection that is not metric compatible. This classification highlights that curvature is a property of the connection and not of the metric tensor or the manifold. It becomes a property of the metric only through the use of the Levi-Civita connection. GR can be equivalently formulated in terms of either of these connections. All of them can be used to define Lagrangians of which Euler-Lagrange equations coincide with the Einstein equations for a particular choice of contributing terms.

 In recent years teleparallel theories have gained more attention as alternative theories of gravity. While one mostly works in the torsion-based setting, there has been interest in the direction of symmetric teleparallelism, where instead of curvature or torsion gravity is effectively described by nonmetricity. It should be mentioned that the mechanism mediating gravity is the affine connection rather than the physical manifold. This is reflected in the fact that in GR curvature is a property of the connection and not of the manifold itself, and thus can be equally described by other connection property such as nonmetricity.  Symmetric teleparallel gravity (STG) offers an interesting geometric interpretation of gravitation besides its formulation in terms of a spacetime metric and Levi-Civita connection or its teleparallel formulation. It describes gravity through a connection which is not metric compatible, however is curvature-free and torsion-free. This connection can be simplified to a partial derivative through the so-called coincident gauge [4, 5]. This geometrically implies that vectors do remain parallel at long distances on a manifold [6]. By demanding that the curvature vanishes and that the connection is torsionless, the remaining gravitational information is encoded in nonmetricity contributions [4, 6-9].

  In STG, the metricity condition of GR is relaxed and then produces teleparallel equivalent of general relativity [10]. One of important properties of STG is their ability to separate gravitational and inertial effects [11] which is not possible in GR. It has produced many strains on the theory such as the issue of defining a gravitational energy-momentum tensor [12].

  More recently, STG was deeply analyzed in [4]. An exceptional class with a vanishing affine connection was discovered. Based on this property, a simpler geometrical formulation of GR was proposed. It leads to a purely inertial connection that can be completely removed by a coordinate gauge choice [13]. This formulation fundamentally deprives gravity from any inertial character. The resulting theory is described by the Hilbert action purged from the boundary term and is more robustly underpinned by the spin-2 field theory. This construction also provides a novel starting point for modified gravity theories, and presents new and simple generalizations where analytical self-accelerating cosmological solutions arise naturally in the early and late time universe.

  Even though the physical aspects of the dynamical nonmetricity are not easy to interpret, a gauge interpretation of STG was proposed from a physical point of view [11]. The equivalence principle which allows eliminating locally gravitation makes it to be an integrable gauge theory. In this approach, the metric represents the gravitational field and the connection corresponds to the gauge potential [14].

  STG was presented originally in a paper by Nester and Yo [6], where the authors emphasize that the formulation brings a new perspective to bear on GR. However, the formulation is geometric and covariant. More recently, some new results and important developments have been obtained. An exceptional class was discovered which is consistent with a vanishing affine connection [7]. Based on this remarkable property, a simpler geometrical formulation of GR was proposed. It provides a novel starting point for modified gravity theories, and presents new and simple generalizations where accelerating cosmological solutions arise naturally in the early and late-time universe.

  In a generalization of STG [15], a nonminimal coupling of a scalar field to the nonmetricity invariant was introduced. The similarities and differences with analogous scalar-curvature and scalar-torsion theories were discussed. The class of scalar-nonmetricity theories was extended by considering a five-parameter quadratic nonmetricity scalar and including a boundary term [16]. The equivalents for GR and ordinary (curvature based) scalar-tensor theories were obtained as particular cases. In a recent paper [5], STG is extended by considering a new class of theories where the nonmetricity is coupled nonminimally to the matter Lagrangian. The theoretical consistency and motivations on this extension are established. Its cosmological application provides a gravitational alternative to dark energy.

  However, comparing to these complicated general approaches mentioned above, a more special and simple model is more suitable to the cosmological application. In this paper, starting from a rather simple Lagrangian a good toy model is obtained, where the accelerating expansion is an intrinsic property of the universe geometry without need of either exotic dark energy or extra fields. This theory is identified as a metrical formulation of $f(T)$ gravity. In contrast with $f(T)$  gravity, this new theory respects local Lorentz symmetry and harbour no extra degrees of freedom, since the dynamic variable is the metric instead of the tetrad. At the same time it has the advantage over $f(R)$ gravity that its field equations are second-order instead of fourth-order and then is free of pathologies. It may provide an satisfactory alternative to conventional dark energy in general relativistic cosmology and a new explanation of the acceleration of the cosmic expansion. The cosmological constant problem, the fine tuning problem and the problem of phantom divide line crossing are solved or disappear.

\section{$\text{ Field equations}$}

The symmetric teleparallel connection relies only on nonmetricity and does not possess neither curvature nor torsion which yields some interesting results. One can transform to a zero connection gauge and thereby covariantize the partial derivatives as well as split the Einstein-Hilbert action into the Einstein Lagrangian density and a boundary term [4, 6].

In differential geometry, the general affine connection can always be decomposed into three independent components [17], namely,
\begin{equation}
\Gamma ^{\lambda }{}_{\mu \nu }=\left\{ ^{\lambda }{}_{\mu \nu }\right\}+K^{\lambda }{}_{\mu \nu }+L^{\lambda }{}_{\mu \nu },\label{A-connection}
\end{equation}
where the first term is the Levi-Civita connection of the metric $g_{\mu\nu }$, given by the standard definition
\begin{equation}
\left\{ _{\mu }{}^{\lambda }{}_{\nu }\right\} \equiv \frac{1}{2}g^{\lambda\beta }\left( \partial _{\mu }g_{\beta \nu }+\partial _{\nu }g_{\beta \mu
}-\partial _{\beta }g_{\mu \nu }\right).
\end{equation}
The second term $K^{\lambda }{}_{\mu \nu }$ is the contortion:
\begin{equation}
K^{\lambda }{}_{\mu \nu }\equiv \frac{1}{2}T^{\lambda }{}_{\mu \nu }+T_{(\mu}{}^{\lambda }{}_{\nu )},
\end{equation}
with the torsion tensor defined as $T^{\lambda }{}_{\mu \nu }$ $\equiv 2\Gamma ^{\lambda }{}_{\left[ \mu \nu \right] }$. The third term is the
disformation
\begin{equation}
L^{\lambda }{}_{\mu \nu }\equiv \frac{1}{2}g^{\lambda \rho }\left( -Q_{\mu\rho \nu }-Q_{\nu \rho \mu }+Q_{\rho \mu \nu }\right),
\end{equation}
which is defined in terms of the nonmetricity tensor: $Q_{\rho\mu\nu }\equiv \nabla _{\rho }g_{\mu\nu }$. We define two traces of the nonmetricity tensor:
\begin{equation}
Q_{\rho }=Q_{\rho }{}^{\mu }{}_{\mu },\widetilde{Q}_{\rho }=Q{}^{\mu}{}_{\rho \mu },
\end{equation}
and introduce the superpotential
\begin{equation}
4P{}^{\alpha }{}_{\mu \nu }=-Q{}^{\alpha }{}_{\mu \nu }+2Q_{(\mu }{}^{\alpha}{}_{\nu )}-Q{}^{\alpha }{}g_{\mu \nu }-\widetilde{Q}{}^{\alpha }{}g_{\mu
\nu }-\delta _{(\mu }^{\alpha }Q_{\nu )}.
\end{equation}
We have therefore
\begin{equation}
L^{\lambda }{}_{\mu \nu }\equiv \frac{1}{2}g^{\lambda \rho }\left( -\nabla_{\mu }g_{\rho \nu }-\nabla _{\nu }g_{\mu \rho }+\nabla _{\rho }g_{\mu \nu
}\right) ,
\end{equation}
and can construct an \textbf{invariant}, the quadratic nonmetricity scalar
\begin{equation}
Q=-g^{\mu \nu }\left( L^{\alpha }{}_{\beta \mu }L^{\beta }{}_{\nu \alpha}-L^{\alpha }{}_{\beta \alpha }L^{\beta }{}_{\mu \nu }\right) ,
\end{equation}
that is special among the general quadratic combination because, in addition to being invariant under local general linear transformations, it is also
invariant under a translational symmetry that allows to completely remove the connection. In this special gauge with vanishing connection, $L^{\lambda
}{}_{\mu \nu }=-\left\{ _{\mu }{}^{\lambda }{}_{\nu }\right\} $, and then the nonmetricity scalar $Q$ can be expressed in terms of the Levi-Civita
connection as
\begin{equation}
Q=-g^{\mu\nu }\left( \left\{ _{\beta }{}^{\alpha }{}_{\mu }\right\} \left\{ _{\nu
}{}^{\beta }{}_{\alpha }\right\} -\left\{ _{\beta }{}^{\alpha }{}_{\alpha
}\right\} \left\{ _{\mu }{}^{\beta }{}_{\nu }\right\} \right) .
\end{equation}
This gauge choice is called the coincident gauge, and shown to be consistent in the symmetric teleparallel geometry [4].

The difference between the invariant $Q$ and the Ricci scalar is a boundary term. The theory described by $Q$  (the boundary term in this theory is absent) is a kind of special STG [4] which is equivalent to an improved version of GR. The connection can be fully trivialised and represents a much simpler geometrical interpretation of gravity, the origins of the tangent space and the spacetime coincide. This theory is called the \emph{coincident GR}, a symmetric teleparallel equivalent of GR.

We start to reformulate GR using the symmetric teleparallel connection and extend it by considering the action defined by a function $f\left( Q\right) $:
\begin{equation}
S=\frac{1}{2\kappa ^{2}}\int d\Omega \left( \sqrt{-g}f\left( Q\right)+L_{m}\right),
\end{equation}
where $\kappa ^{2}=8\pi G_{N}$ with the bare gravitational constant $G_{N}$, $L_{m}$ is the Lagrangian of the matter fields. The variational principle
yields the field equations for the metric $g_{\mu \nu }$:
\begin{equation}
\frac{2}{\sqrt{-g}}\nabla _{\alpha }\left( \sqrt{-g}f^{\prime }P{}^{\alpha }{}_{\mu \nu }\right) +\frac{1}{2}g_{\mu \nu }f+f^{\prime }\left( P_{\mu
\alpha \beta }Q_{\nu }{}^{\alpha \beta }-2Q_{\alpha \beta \mu }P^{\alpha \beta }{}_{\nu }\right) =-\kappa ^{2}T_{\mu \nu }.
\end{equation}
where primes ($\prime $) stand for derivatives of the functions with respect to $Q$, and
\begin{equation}
T_{\rho \sigma }=-\frac{2}{\sqrt{-g}}\frac{\delta L_{m}}{\delta g^{\rho\sigma }},
\end{equation}
is the energy-momentum of the matter fields.

\section{$\text{Cosmological model}$}
We now investigate the cosmological dynamics for the models based on $f(Q)$ gravity. In order to derive conditions for the cosmological viability of $f(Q)$ models we shall carry out a general analysis without specifying the form of $f(Q)$ at first. Consider a flat Friedmann-Lemaitre-Robertson-Walker (FLRW) background with the metric
\begin{equation}
g_{\mu \nu }=\text{diag}\left( -1,a\left( t\right) ^{2},a\left( t\right)^{2},a\left( t\right) ^{2}\right),
\end{equation}
where $a(t)$ is a scale factor. The non-vanishing components of the Levi-Civita connection are
\begin{eqnarray}
\left\{ _{0}{}^{0}{}_{0}\right\}  &=&0,\left\{ _{0}{}^{0}{}_{i}\right\}=\left\{ _{i}{}^{0}{}_{0}\right\} =0,\left\{ _{i}{}^{0}{}_{j}\right\} =a
\overset{\cdot }{a}\delta _{ij},\nonumber \\
\left\{ _{0}{}^{i}{}_{0}\right\}  &=&0,\left\{ _{j}{}^{i}{}_{0}\right\}=\left\{ _{0}{}^{i}{}_{j}\right\} =H\delta _{j}^{i},\left\{
_{j}{}^{i}{}_{k}\right\} =0,i,j,k,...=1,2,3.
\end{eqnarray}
and then the nonmetricity scalar is
\begin{equation}
Q=-6H^{2}.
\end{equation}
Here $H\equiv \overset{\cdot }{a}/a$ is the Hubble parameter and a dot represents a derivative with respect to the cosmic time $t$. Then the field
equations (11) take the forms
\begin{equation}
3H^{2}=-\frac{1}{2f^{\prime }}\left( f+6H^{2}f^{\prime }\right) -18H^{2} \overset{\cdot }{H}\frac{f^{\prime \prime }}{f^{\prime }}+\frac{\kappa ^{2}}{f^{\prime }}\rho ,
\end{equation}
\begin{equation}
-2\overset{\cdot }{H}-3H^{2}=\frac{1}{2f^{\prime }}\left( f+6H^{2}f^{\prime}\right) -18H^{2}\overset{\cdot }{H}\frac{f^{\prime \prime }}{f^{\prime }}+
\frac{\kappa ^{2}}{f^{\prime }}p,
\end{equation}
which lead to
\begin{equation}
\overset{\cdot }{H}=-\frac{\kappa ^{2}}{2f^{\prime }}\left( \rho +p-\frac{36}{\kappa ^{2}}H^{2}\overset{\cdot }{H}f^{\prime \prime }\right) ,
\end{equation}
where $\rho =-T^{0}{}_{0}$, and $p=T^{1}{}_{1}=T^{2}{}_{2}=T^{3}{}_{3}$ are the density and the pressure of the matter fields, respectively. It is easy
to see that, in $f(Q)$ gravity, the gravitational constant $\kappa ^{2}$ is replaced by an effective (time dependent) $\kappa _{\text{eff}}^{2}=\kappa
^{2}/f^{\prime }(Q)$. On the other hand, it is reasonable to assume that the present day value of $\kappa _{\text{eff}}^{2}$ is the same as the $\kappa
^{2}$ so that we get the simple constraint :
\begin{equation}
\kappa _{\text{eff}}^{2}(z=0)=\kappa ^{2}\rightarrow f^{\prime}(Q_{0})=1,
\end{equation}
where $z$ is the redshift.

If $f\left( Q\right) $ has the form
\begin{equation}
f\left( Q\right) =Q+\Phi \left( Q\right) ,
\end{equation}
then the equations (16), (17) and (18) become
\begin{equation}
3H^{2}=\kappa ^{2}\left( \rho +\rho _{\text{de}}\right) ,
\end{equation}
\begin{equation}
-2\overset{\cdot }{H}-3H^{2}=\kappa ^{2}\left( p+p_{\text{de}}\right),
\end{equation}
and
\begin{equation}
\frac{\overset{\cdot \cdot }{a}}{a}=-\frac{\kappa ^{2}}{6}\left( \rho +\rho_{\text{de}}+3p+3p_{\text{de}}\right) ,
\end{equation}
where
\begin{equation}
\rho _{\text{de}}=-\frac{1}{2\kappa ^{2}}\Phi -\frac{6}{\kappa ^{2}}H^{2}\Phi ^{\prime }-\frac{18}{\kappa ^{2}}H^{2}\overset{\cdot }{H}\Phi
^{\prime \prime },
\end{equation}
and
\begin{equation}
p_{\text{de}}=\frac{1}{2\kappa ^{2}}\Phi +\frac{6}{\kappa ^{2}}H^{2}\Phi^{\prime }+\frac{2}{\kappa ^{2}}\overset{\cdot }{H}\left( \Phi ^{\prime
}-9H^{2}\Phi ^{\prime \prime }\right) ,
\end{equation}
are the density and the pressure of the "geometrical dark energy" $\Phi \left( Q\right) $, respectively. The state
equation of the "geometrical dark energy" is then
\begin{equation}
w_{\text{de}}=\frac{p_{\text{de}}}{\rho _{\text{de}}}=-1-\frac{2\overset{\cdot }{H}\left( \Phi ^{\prime }-18H^{2}\Phi ^{\prime \prime }\right) }{\frac{1}{2}\Phi +6H^{2}\Phi ^{\prime }+18H^{2}\overset{\cdot }{H}\Phi^{\prime \prime }}.
\end{equation}

Let $N=\ln a$. For any function $F(a)$ we have $F_{N}=:\frac{dF}{dN}=H\overset{\cdot }{F}.$ Then the equation (26) becomes
\begin{equation}
w_{\text{de}}=-1+\frac{1}{3}\frac{Q_{N}}{Q}\frac{\Phi ^{\prime }+3Q\Phi^{\prime \prime }}{\Phi /Q-2\Phi ^{\prime }+\frac{1}{2}Q_{N}\Phi ^{\prime\prime }}.
\end{equation}
The equations (21) and (22) become
\begin{equation}
H^{2}=\frac{\kappa ^{2}}{3}\rho -\frac{1}{6}\Phi +\frac{1}{3}Q\Phi ^{\prime}+Q\overset{\cdot }{H}\Phi ^{\prime \prime },
\end{equation}
\begin{equation}
\left( H^{2}\right) _{N}=-\frac{2\kappa ^{2}p-Q+\Phi -2Q\Phi ^{\prime }}{3Q\Phi ^{\prime \prime }+2\Phi ^{\prime }+2}.
\end{equation}
We see that a constant $\Phi $ acts just like a cosmological constant (dark energy), and $\Phi $ linear in $Q$ (i.e. $\Phi ^{\prime }=$ constant) is
simply a redefinition of gravitational constant $\kappa ^{2}$. Then the equation (29) can be written as
\begin{equation}
\frac{1}{6}\left( 1+\Phi ^{\prime }-\frac{3}{2}Q\Phi ^{\prime \prime}\right) Q_{N}=-\frac{Q}{2}+\frac{1}{2}\Phi -Q\Phi ^{\prime }+\kappa^{2}p.
\end{equation}
Taking a universe with only dust matter so
\begin{equation}
p=0,\nonumber
\end{equation}
we have
\begin{equation}
-\frac{1+\Phi ^{\prime }-\frac{3}{2}Q\Phi ^{\prime \prime }}{3Q\left( 1-\Phi/Q+2\Phi ^{\prime }\right) }dQ=dN,
\end{equation}
and then we find the solution $Q(a)$ in closed form:
\begin{equation}
a\left( Q\right) =\exp \left\{ -\frac{1}{3}\int_{-6H_{0}^{2}}^{Q}\frac{dx}{x}\frac{1+\Phi _{x}\left( x\right) -\frac{3}{2}x\Phi _{xx}\left( x\right) }{
1-\Phi \left( x\right) /x+2\Phi _{x}\left( x\right) }\right\} .
\end{equation}

The equations (28), (29) and (32) take the same forms as the ones for $f(T)$ gravity given by Linder [18]. In a sense our model can be considered as a metric formulation of the $f(T)$ model.

The equations (28) and (29) become
\begin{equation}
3H^{2}=-\frac{1}{2}\Phi -6H^{2}\Phi ^{\prime }-18H^{2}\overset{\cdot }{H}\Phi ^{\prime \prime }+\kappa ^{2}\rho .
\end{equation}
\begin{equation}
-2\overset{\cdot }{H}-3H^{2}=\frac{1}{2}\Phi +6H^{2}\Phi ^{\prime }+2\overset{\cdot }{H}\left( \Phi ^{\prime }-9H^{2}\Phi ^{\prime \prime }\right)+\kappa ^{2}p.
\end{equation}
and yield
\begin{equation}
\overset{\cdot }{H}=-\frac{\kappa ^{2}}{2\left( 1+\Phi ^{\prime}-18H^{2}\Phi ^{\prime \prime }\right) }\left( \rho +p\right) .
\end{equation}
So, $\frac{\kappa ^{2}}{1+\Phi ^{\prime }-18H^{2}\Phi ^{\prime \prime }}$ is simply a redefinition of gravitational constant $\kappa ^{2}$.

Considering a vacuum  universe with the density and pressure of universal matter: $\rho=0$ and $p=0$, Eq. (35) gives a $\textbf{universal}$ de Sitter solution
\begin{equation}
\overset{\cdot }{H}=0,
\end{equation}
for \emph{any function} $\Phi \left( Q\right) $. Then (26) gives
\begin{equation}
w_{\text{de}}=-1,
\end{equation}
which means that the role of the geometrical quantity $\Phi(Q)$ can be seen as the "dark energy" or the cosmological constant.

The equation (34) gives
\begin{equation}
\overset{\cdot }{H}=-\frac{\kappa ^{2}p+3H^{2}+\frac{1}{2}\Phi +6H^{2}\Phi ^{\prime }}{2\left( 1+\Phi ^{\prime }-9H^{2}\Phi ^{\prime \prime }\right) }.
\end{equation}

Substituting (38) into (26) yields
\begin{equation}
w_{\text{de}}=-\frac{\Phi +6H^{2}\Phi ^{\prime }+54H^{4}\Phi ^{\prime \prime}+2\kappa ^{2}\left( 9H^{2}\Phi ^{\prime \prime }-\Phi ^{\prime }\right) p}{
\left( 1+\Phi ^{\prime }\right) \left( \Phi +12H^{2}\Phi ^{\prime }\right)-18H^{2}\Phi ^{\prime \prime }\left( 12H^{2}\Phi ^{\prime }+3H^{2}+\Phi
\right) -18H^{2}\Phi ^{\prime \prime }\kappa ^{2}p}.
\end{equation}

According to the Eqs. (16-18) and (26), we can see that the evolutions of cosmological quantities depend on the choice of $f(Q)$, its first and second derivatives with respect to $Q$.   As an example, we take a concrete form as
\begin{equation}
\Phi \left( Q\right) =\alpha \left( -Q\right) ^{n}=\alpha 6^{n}H^{2n}.
\end{equation}
In this case, (38) and (39) become
\begin{equation}
\overset{\cdot }{H}=-\frac{\kappa ^{2}p+3H^{2}+\left( \frac{1}{2}-n\right)\alpha 6^{n}H^{2n}}{2\left( 1+3n\left( 3n-5\right) \alpha6^{n-2}H^{2n-2}\right) },
\end{equation}
and
\begin{equation}
w_{de}=-\frac{3\left( 3n-2\right) \left( n-1\right) H^{2}+n\left(3n-1\right) \kappa ^{2}p}{-3\left( n+1\right) \left( 3n-2\right)
H^{2}+\allowbreak 6^{n}n\left( 2n-1\right) \left( 3n-2\right) \alpha H^{2n}\allowbreak -3n\left( n-1\right) \kappa ^{2}p}.
\end{equation}

For dust matter
\begin{equation}
p=0,
\end{equation}
we have
\begin{equation}
\overset{\cdot }{H}=-\frac{3H^{2}+\left( \frac{1}{2}-n\right) \alpha6^{n}H^{2n}}{2\left( 1+3n\left( 3n-5\right) \alpha 6^{n-2}H^{2n-2}\right) },
\end{equation}
\begin{equation}
w_{\text{de}}=-\frac{3\left( 3n-2\right) \left( n-1\right) H^{2}}{-3\left(n+1\right) \left( 3n-2\right) H^{2}+\allowbreak 6^{n}n\left( 2n-1\right)
\left( 3n-2\right) \alpha H^{2n}\allowbreak }.
\end{equation}

When
\begin{equation}
n=2,
\end{equation}
i.e. for the gravitational Lagrangian
\begin{equation}
L_{g}=\sqrt{-g}\left( Q+\alpha Q^{2}\right) ,
\end{equation}
(39) gives
\begin{equation}
w_{\text{de}}=-\frac{1}{3\left( 24\alpha H^{2}-1\right) }.
\end{equation}
Next, we estimate the value of the model parameter $\alpha$.  Letting
\begin{equation}
w_{\text{de}}=-1,\nonumber
\end{equation}
and
\begin{equation}
H_{0}=74km/sec/Mpc\simeq 2.4\times 10^{-18}sec^{-1},
\end{equation}
we can compute to gain
\begin{equation}
\alpha =1.\,0145\times 10^{-5}\left( km/sec/Mpc\right) ^{-2}=9.\,6451\times10^{33}sec^{2}.
\end{equation}
Then when
\begin{equation}
w_{\text{de}}=\emph{\ }-\frac{1}{3},
\end{equation}
we obtain
\begin{equation}
H=90.\,631km/sec/Mpc.
\end{equation}
Observations of type Ia supernovae at moderately large redshifts ($z\sim 0.5$ to $1$) have led to the conclusion that the Hubble expansion of the universe is accelerating [19-22]. This is consistent also with microwave background measurements [23,24]. According to the result of [25-27], $H=90.\,631km/sec/Mpc$ corresponds to
\begin{equation}
z\sim 0.88,
\end{equation}
which is consistent with the observations.

The equation (44) now becomes
\begin{equation}
\overset{\cdot }{H}=\frac{3H^{2}\left( 18\alpha H^{2}-1\right) }{2\left(6\alpha H^{2}+1\right) }.
\end{equation}
\ Using the formula
\begin{equation}
H=\frac{\overset{\cdot }{a}}{a}=-\frac{1}{1+z}\overset{\cdot }{z},
\end{equation}
(54) can be rewritten as
\begin{equation}
-\frac{2\left( 6\alpha H^{2}+1\right) dH}{3H\left( 18\alpha H^{2}-1\right) }=\frac{dz}{\left( 1+z\right) },
\end{equation}
and integrated
\begin{equation}
z=\left( \frac{H^{2}}{H_{0}^{2}}\right) ^{1/3}\left( \frac{18\alpha H_{0}^{2}-1}{18\alpha H^{2}-1}\right) ^{4/9}-1.
\end{equation}

The equation (21) indicates that during the evolution of the universe $H^{2}$ decreases owing to decreasing of the matter density $\rho $. This makes $w_{\text{de}}$ descend during the evolution of the universe. The evolution of the function $w_{de}=w_{de}(\alpha H^{2})$ given by (48) is illustrated in Fig. 1. It is shown that the state equation of  "geometrical dark energy" can cross over the phantom divide line ($w_{de}=-1$). According to (48) when $H^{2}=\frac{1}{12\alpha }$, $w_{\text{de}}=$ $-\frac{1}{3}$, $\Phi
\left( Q\right) $ changes from \textquotedblright visible\textquotedblright\ to dark as indicated by (23). If $H^{2}>\frac{1}{12\alpha }$, it decelerates
the expansion, if $H^{2}<\frac{1}{12\alpha }$, it accelerates the expansion. When $H^{2}=\frac{1}{18\alpha }$, $w_{\text{de}}$ crosses the phantom divide
line $-1$. In other words, the expansion of the universe naturally includes a decelerating and an accelerating phase. $\alpha $ given by (50) can be
seen as a new constant describing the evolution of the universe.

\begin{figure}[ht]
  \includegraphics[width=7cm]{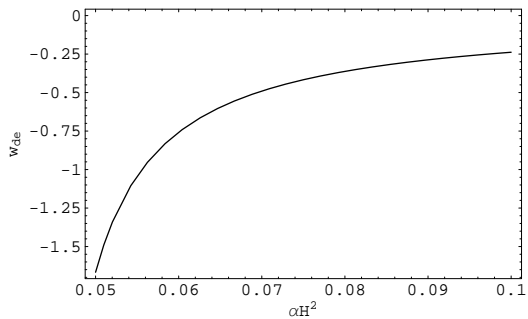}\\
  \caption{ The evolution of $w_{de}$.}\label{figure-wde}
\end{figure}

\section{$\text{Dynamical system approach in the STG theory}$}
 A powerful and elegant  way to investigate the universal dynamics is to recast the cosmological equations into a dynamical system \cite{ds-1807}. Dynamical system approach can be applied to analyze the stability of system, and has been  studied in the plenty of cosmological models, such as the canonical scalar-field models \cite{ds-canonical1,ds-canonical2}, the non-canonical scalar-field models \cite{ds-noncanonical1,ds-noncanonical2,ds-noncanonical3},  the scalar-tensor theories \cite{ds-scalar-tensor1,ds-scalar-tensor2}, the $f(R)$ gravity theory \cite{ds-fr1,ds-fr2,ds-fr3}, etc \cite{ds-refs1,ds-refs2,ds-refs3,ds-refs4,ds-refs5}. For the recent review on the dynamical system  approach, one can see reference \cite{ds-review}. In this section, we study the cosmological dynamical system in the STG theory. Following the above discussion, we consider a specific form, i.e. Eq. (47). Then the cosmological equations (16,18) can be rewritten as
\begin{equation}
3H^{2}=-\frac{36\dot{H}H^{2}\alpha}{1-12H^{2}\alpha}
-\frac{-6H^{2}+36H^{4}\alpha+6H^{2}(1-12H^{2}\alpha)}{2(1-12H^{2}\alpha)}+\frac{\kappa\rho}{1-12H^{2}\alpha}\label{eq-h2-concrete}
\end{equation}
\begin{equation}
\dot{H}=\frac{\kappa\rho+\kappa p}{2(-1+48H^{2}\alpha)}.\label{eq-doth-concrete}
\end{equation}
We assume that the universal matter include the dust matter and the radiation matter. Combining Eqs. (\ref{eq-h2-concrete}) and (\ref{eq-doth-concrete}), we gain
\begin{equation}
\frac{6H^{2}\alpha}{1-12H^{2}\alpha}+
\frac{\kappa\rho_{r}(1-24H^{2}\alpha)}{3H^{2}(1-12H^{2}\alpha)(1-48H^{2}\alpha)}
+\frac{\kappa\rho_{m}(1-30H^{2}\alpha)}{3H^{2}(1-12H^{2}\alpha)(1-48H^{2}\alpha)}=1,\label{eq-friedmann-concrete}
\end{equation}
where $\rho_{m}$ and $\rho_{r}$  denote the density of dust matter and radiation matter, respectively.  The relations between pressure and density for two matters:   $p_{m}=0$ and $p_{r}=\frac{1}{3}\rho_{r}$ have been used in the above derivation. Defining the following three dimensionless variables
\begin{equation}
x=\frac{6H^{2}\alpha}{1-12H^{2}\alpha},
~~~~y=\frac{\kappa\rho_{r}(1-24H^{2}\alpha)}{3H^{2}(1-12H^{2}\alpha)(1-48H^{2}\alpha)},
~~~~\tilde{\Omega}_{m}=\frac{\kappa\rho_{m}(1-30H^{2}\alpha)}{3H^{2}(1-12H^{2}\alpha)(1-48H^{2}\alpha)},\label{eq-dimensionless-variables}
\end{equation}
 then Eq. (\ref{eq-friedmann-concrete}) can be rewritten as
\begin{equation}
\tilde{\Omega}_{m}=1-x-y,\label{eq-dimensionless-constraint}
\end{equation}
which can be seen as a constraint equation.

The dynamical system in the STG is obtained as follows
\begin{equation}
x_{N}=-\frac{x(1+2x)[3+6x^{2}+y-3x(3+2y)]}{1-5x+6x^{2}},\label{eq-xp}
\end{equation}
\begin{equation}
y_{N}=\frac{y[-1+x-144x^{5}+y-20xy-48x^{3}(-2+3y)+48x^{4}(1+3y)+8x^{2}(-5+13y)]}{(1-2x)^{2}(1-9x+18x^{2})},\label{eq-yp}
\end{equation}
where subscript "N" denotes the derivative with respect to $N=\ln a$.
Solving the equations $x_{N}=0$ and $y_{N}=0$, we can get five critical points of the dynamical system for the STG theory, which are listed in Table \ref{table-DS}.   Next, we investigate the stability of critical points.   Taking $x_{j}=x_{cj}+\delta x_{j}$, we can linearize the dynamical equations to obtain $\delta x^{'}_{j}=\cal{M}$$\delta x_{j}$,  where $x_{j}$ denote the dimensionless variable $(x,y)$ for $j=1$ and $j=2$ respectively, $x_{cj}$ denote critical points, $\delta x_{j}$ denote the linear perturbations of the dynamical variables, and $\cal{M}$ denotes the coefficients matrix. It is well known that the eigenvalues $\lambda_{i}$ of $\cal{M}$ determine the stability of critical points, i.e. the stability of critical points relate to the symbols of real parts of $\lambda_{i}$. The critical point is unstable, if  its corresponding eigenvalues have  one or more positive real part; While the critical point is stable, if  its corresponding eigenvalues have all the  negative real part. Solving the secular equation $|\cal{M}$-$\lambda E|=0$ ($E$ is the unit matrix), we can obtain the eigenvalues of the critical points (please see Table \ref{table-DS}). From Table \ref{table-DS}, we can see that the critical points: $P_{4}$ and  $P_{5}$ are stable. We also calculate the values of the dimensionless variable $\tilde{\Omega}_{m}$ and the effective state parameter of total matter in our universe $w_{tot}^{eff}=-1-\frac{2\dot{H}}{3H^{2}}=\frac{6x-12x^2-6xy+y}{3-15x+18x^2}$ for each critical point. After calculation, it is shown that stable critical point $P_{4}$ corresponds to the geometrical dark energy dominated de Sitter universe  ($w_{tot}^{eff}=-1$), while $P_{5}$ corresponds to the matter dominated universe.

\begin{figure}[ht]
  \includegraphics[width=8cm]{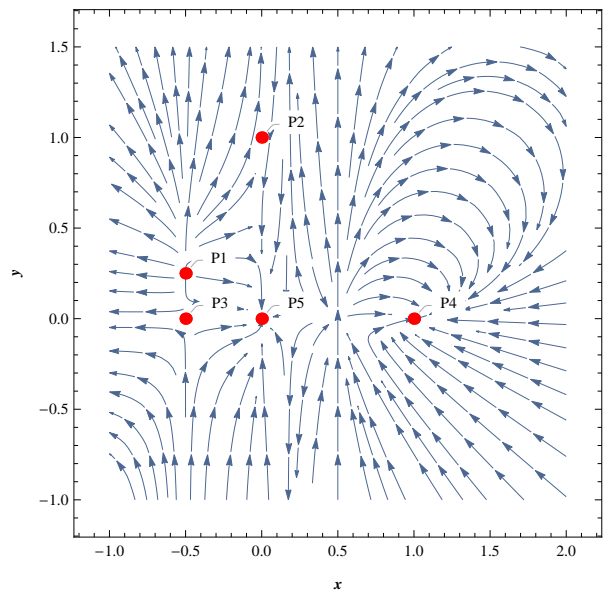}\\
  \caption{Phase-space trajectories on the $x-y$ plane for STG. The red dot corresponds to the critical point. }\label{figure-ps}
\end{figure}

We plot the phase space portraits  on the $x-y$ plane for STG, which are illustrated in Fig.\ref{figure-ps}. From Fig.\ref{figure-ps}, we can read that
 $P_{4}$ and $P_{5}$ represent  attractor solutions of the dynamical system, while there are not attractor solutions corresponding to $P_{1}$, $P_{2}$ and $P_{3}$ critical points.  Given that $P_{4}$ represents an attractors for the universe at late times (at where surrounding points are attracted into it irrespective of the initial conditions), it is shown that the  cosmological constant problems, such as the fine tuning problem, could be solved in this STG model.

\begin{table}[!htbp]
 \vspace*{-12pt}
 \begin{center}
 \begin{tabular}{ |c| c| c|c| c | c | c |} \hline\hline
    Points  & ~~~$x$~~~    & ~~~$y$~~~   & ~~$\tilde{\Omega}_{m}$~~  & ~~$w_{tot}^{eff}$~~    &   ~~Eigenvalues~~        &   ~~Stability~~     \\\hline
    $P_{1}$ & $-\frac{1}{2}$ & $\frac{1}{4}$ & $\frac{5}{4}$             &$-\frac{1}{3}$  &   $(2,\frac{2}{5})$             & Unstable  \\\hline
    $P_{2}$ & 0 & 1                     &  $0$                      &$\frac{1}{3}$   &   $(-4,1)$                      &  Unstable \\\hline
    $P_{3}$ & $-\frac{1}{2}$  & 0          &  $\frac{3}{2}$            &$-\frac{2}{5}$  &   $(\frac{9}{5},-\frac{2}{5})$  & Unstable  \\\hline
    $P_{4}$ & 1   & 0                       &  $0$                      &$-1$            &   $(-\frac{9}{2},-4)$           & Stable  \\\hline
    $P_{5}$ & 0  &  0                       &  $\frac{5}{4}$            &$0$             &   $(-3,-1)$                     & Stable    \\\hline\hline
 \end{tabular}
 \end{center}
 \caption{The stability and eigenvalues of critical points for STG theory.}\label{table-DS}
 \end{table}

\section{$\text{Discussion and conclusions}$}

A cosmology of Symmetric teleparallel gravity has been developed. We focus on the case of the extended Coincident General Relativity [4]. The gravitational field equation and the cosmological equations are derived. It gives a explaination to the acceleration of the cosmic expansion. It is found that the role of dark energy can be played by the geometry itself and then is endowed with intrinsic character of the spacetime.  The dark energy is identified with the geometry of the spacetime. Now we are returning to the original idea of Einstein and Wheeler: gravity is a geometry [47, 48]. The analytic expressions of the density and the pressure of the geometric dark energy, their state equation and the density parameters are derived. The main results are equations (21)-(26). Eqs. (21) and (22) correspond to the Friedmann equation and the Raychaudhuri equation respectively, while (23) is the acceleration equation, which represent the Einstein frame of the theory. It is easy to see that when $f(Q)=Q$, they reduce to the Friedmann cosmology. It should be noted that although (21) and (22) have the same form as the Friedmann equations, the solutions (28) and (35) are different. The reason is that in (21) and (22) the density and the pressure of the geometrical dark energy  are functions of geometry as indicated by (24) an (25). This is a geometrical dynamic model of dark energy and thus is different from the $\Lambda$CDM model essentially. As shown in  equation (26) or (48), the state equation of the geometrical dark energy can cross over the phantom divide line ($w_{de}= -1$) in the STG model.

Furthermore, the dynamical system method are investigated in the STG theory. It is shown that there are five critical points in the STG model for taking $f(Q)=Q+\alpha Q^2$. The critical points $P_{4}$ and $P_{5}$ are stable.  $P_{4}$ corresponds to the geometrical dark energy dominated de Sitter universe ($w_{tot}^{eff}$=-1), while $P_{5}$ corresponds to the matter dominated universe ($w_{tot}^{eff}$=0). Given that $P_{4}$ is an attractor, the  cosmological constant problems, such as the fine tuning problem, could be solved in the STG model.

\textbf{\ Acknowledgments }
We thank the anonymous referee for his/her very instructive comments, which improve our paper greatly. The research work is supported by   the National Natural Science Foundation of China (11645003,11705079).

\end{document}